\documentclass[%
nofootinbib,
reprint,
superscriptaddress,
showpacs,
 amsmath,amssymb,
prd,
]{revtex4-1}

\usepackage{graphicx}% Include figure files
\usepackage{dcolumn}% Align table columns on decimal point
\usepackage{bm}% bold math
\usepackage{color}

\usepackage[pagewise]{lineno}
\AtBeginDocument{%
  \setlength{\linenumbersep}{\dimexpr\oddsidemargin+0.5in-\linenumberwidth\relax}%
}

\newcommand{\dl}{DL}

\usepackage{amssymb} %% <- for \square and \blacksquare
\newcommand*{\QEDA}{\hspace{9cm}\null\nobreak\hfill\ensuremath{\blacksquare}}%

\begin{document}

%\preprint{APS/123-QED}

%\title{Quantization via Exactly Resummed Wentzel-Kramers-Brillouin Series}
\title{Resummed Wentzel-Kramers-Brillouin Series: Quantization and Physical Interpretation}
% Force line breaks with \\

\author{B.~Tripathi}
\affiliation{Department of Physics, University of Wisconsin-Madison, Madison, Wisconsin 53706, USA}
\email{btripathi@wisc.edu}

\date{\today}% It is always \today, today,
\begin{abstract}
The Wentzel-Kramers-Brillouin (WKB) perturbative series, a widely used technique for solving linear waves, is typically divergent and at best, asymptotic, thus impeding predictions beyond the first few leading-order effects. Here, we report a closed-form formula that exactly resums the perturbative WKB series to all-orders for two turning point problem. The formula is elegantly interpreted as the action evaluated using the product of spatially-varying wavenumber and a coefficient related to the wave transmissivity; unit transmissivity yields the Bohr-Sommerfeld quantization. 

\end{abstract}

%\begin{description}
% Secondary publications and information retrieval purposes.
% \item[PACS numbers]
% May be entered using the \verb+\pacs{#1}+ command.
% You may use the \texttt{description} environment to structure your abstract;
% use the optional argument of the \verb+\item+ command to give the category of each item. 
%\end{description}

%\pacs{96.60.qd, 96.60.Hv}% PACS, the Physics and Astronomy
\maketitle

%\tableofcontents
\section{Introduction} 

Linear waves are ubiquitous in the world of physics with applications ranging from quantum mechanics \cite{berry1972} to electromagnetism, fluid dynamics, and astrophysics \cite{iyer1987}. The properties of these waves are encoded in their dispersion relations, which reveal the nature of the medium they traverse, as well as their generating source \cite{iyer1987,fuller2015} (e.g., gravitational waves). This problem of obtaining the dispersion relation is traditionally addressed with the Wentzel-Kramers-Brillouin (WKB) perturbative series \cite{benderorszag}, which however is typically divergent and at best, asymptotic \cite{berry1972, robnik1997}. It thus presents challenges to predict phenomena beyond the leading-order effects. Obtaining an expression for this series to all-orders in perturbation theory, preferably in a closed-form, is therefore desirable. Despite its usefulness in finding previously unknown physical interpretations of fully quantized wave, it has remained elusive. Here, we will be accomplish both the tasks of finding a closed-form formula and assigning a physical meaning to it.

In the quest for developing a closed-form quantization condition (dispersion relation) for linear waves, several insightful but hitherto-unsuccessful attempts have been undertaken through various ways like investigating structures of higher order expressions in the WKB series  \cite{bender1977, barclay1994}, utilization of supersymmetric WKB method \cite{susy1985,susy1995}, complex WKB method \cite{voros1983}, and phase integral method \cite{froman1965}. We present here, a simple and insightful method to achieve this aim successfully.

The unexpected simplicity (both in mathematical structure and geometric-optical interpretation) of the closed-form formula reported here, arising out of the unwieldy WKB series, is what we believe to be the most striking about this work.

Our principal result is that the one-dimensional wave equation\footnote{
The book-keeping parameter $\epsilon$ can be set equal to $1$ at the outset or at the end of the perturbative calculations.} (or Schr\"{o}dinger equation)
\begin{equation} \label{eq:WKB_diff}
\epsilon^2 \frac{d^2 \psi(x)}{dx^2} = Q\left(x\right) \psi(x),\hspace{1.3cm} \psi\left(\pm \infty \right)=0,
\end{equation}
 with $Q\left(x\right) = - k^2(x) = 2m[V(z)-E]/\hslash^2$ where $k(z)$ is the local wavenumber, $V(z)$ is the potential, $E$ is the energy eigenvalue, $m$ and $\hslash$ are the mass and reduced Planck's constant respectively, for the case of two turning points [locations where $Q(z)=0$ with $z$ being a complex variable] has an exact closed-form quantization condition

\begin{equation} \label{eq:quantpot}
\oint_{\Gamma} k(z) \cdot \tau(z) dz = \left(K +\frac{1}{2}\right) 2\pi, \hspace{0.8cm} (K = 0,1,2,...),
\end{equation}
where $T(z)=\tau^2(z)$ is the wave-traversing medium's transmissivity of a layer of width $1/k(z)$ given as \cite{bremmer1951}
\begin{equation} \label{eq:transmissivitydefn}
\tau(z) = \sqrt{1 -\left(\frac{1}{2} \frac{d (k^{-1})}{dz}\right)^{2}} = \sqrt{1+ \left(\frac{S_{1}'}{S_{0}'}\right)^{2}}.
\end{equation}
The contour $\Gamma$ encircles the two turning points in anticlockwise direction. ($S_{0}'$ and $S_{1}'$ are explained immediately below, but presented here due to their elegant appearances.) Unit transmissivity reproduces the commonly known leading-order WKB approximation.

\section{Conventional WKB} 
To begin with, consider the traditional transformation that is applied to the Schr\"{o}dinger equation \eqref{eq:WKB_diff},
\begin{subequations} 
\begin{align} \label{eq:sprime}
    \psi(z,\epsilon)&=\exp\left[ \frac{1}{\epsilon} S(z,\epsilon) \right],\\
    \mathrm{or,\ } S'(z,\epsilon)&=\frac{\epsilon}{\psi(z,\epsilon)} \frac{d\psi(z,\epsilon)}{dz},
\end{align}
\end{subequations}
to obtain the Riccati equation: $\left(S'\right)^2 + \epsilon S^{''}=Q(z)$.

The quantization condition for energy eigenvalue $E$ in Eq.~\eqref{eq:WKB_diff} is given in terms of the WKB eigenfunction's exponent in Eq.~\eqref{eq:sprime} as \footnote{
One way to derive this is by integrating $dS/dz = S'(z,\epsilon)$ in Eq.~\eqref{eq:sprime} along a contour $\Gamma$ such that it encloses, for the $K^{\mathrm{th}}$ energy level, all $K$ zeros of the eigenfunction on the real axis, between the classical turning points. This leads to
$\oint_\Gamma S'(z,\epsilon) dz = \oint_\Gamma \epsilon \frac{d\mathrm{\ ln}\psi(z)}{dz} dz= \epsilon \mathrm{\ ln}\psi(z) \Big|_{\mathrm{evaluated\ once\ around\ }\Gamma}= K \cdot 2 \pi i \epsilon$.}:
\begin{equation} \label{eq:logintegrated}
    \frac{1}{\epsilon} \oint_\Gamma S'(z,\epsilon) dz = K \cdot 2 \pi i.
\end{equation}

This equation, although exact, is not useful unless we know what $S'(z,\epsilon)$ is [or be able to solve the above Riccati equation or Eq.~\eqref{eq:WKB_diff} exactly]. We, therefore, proceed with the perturbative method to compute $S'(z,\epsilon)$ as
\begin{equation} \label{eq:pertWKB}
S'(z,\epsilon) =  \sum_{n=0}^{\infty} \epsilon^n S_{n}'(z).
\end{equation}
Substituting this ansatz in the aforementioned Riccati equation and equating like-powers of $\epsilon$, one finds the $S_{n}'(z)$ to obey the recurrence relation,
\begin{align} \label{eq:recur}
 &S_{0}'(z) =  \sqrt{ Q(z)} = i k(z) ,\\ \label{eq:recurS1}
 &S_{1}'(z) =  -\frac{1}{2} \frac{d}{dz} \mathrm{ln} [S_0^{'}(z)],\\  \label{eq:recur2}
&2 S_{0}' S_{n}' + \sum_{j = 1}^{n-1} S_{j}' S_{n-j}' + S_{n-1}'' = 0, \hspace{0.8cm} (n \geq 2 ).
\end{align}  
Eq.~\eqref{eq:logintegrated} thus, to all-orders in perturbation theory, is (it was first written in this form by Dunham \cite{dunham1932}):
\begin{equation} \label{eq:Dunham1}
\frac{1}{2 i \epsilon} \oint_{\Gamma} \sum_{n=0}^{\infty} \epsilon^n S_n'\left( z \right) dz = K \pi, \hspace{1cm} (K = 0,1,2,...).
\end{equation}

This series on the left hand side (LHS) is now to be summed up. However, it is typically a divergent asymptotic series \cite{brezin1977, stone1978}. One way to circumvent this challenge is to employ the Borel summation technique and assign a physical meaning to such a series. When employed, the analytic continuation of the Borel transform, however, presents another difficulty -- singularities on the integration contour (see, e.g., \cite{zj1981, zj1984, unsal2012, aniceto2015}). Avoiding such with contour deformation yields ambiguous imaginary terms that plague the energy eigenvalues $E$. Significant progresses have been made to address such problems and more recently via exact WKB and uniform WKB methods, following advances of resurgence theory, developed by Ecalle and others in the 1980s \cite{ecalle1981, voros1983, sueishi2020, dunnetransseries2014}. In such works, the ambiguous imaginary terms arising from the Borel summation are made to cancel each other systematically to all-orders by considering a ``resurgent trans-series" for the energy eigenvalues \cite{dunnetransseries2014}, as opposed to a perturbative series for it as we have done here. Such a path although very insightful and useful will not be pursued here as we wish to present an alternative simpler way to resum the diverging series of Eq.~\eqref{eq:Dunham1}, for a class of potentials, and assign a physical meaning to the resummed series.

Note that, in Eq.~\eqref{eq:Dunham1}, the term of first order in $\epsilon$ [i.e., $S'_1(z)$] can be integrated exactly \cite{benderorszag}:
\begin{equation} \label{eq:piover2}
\begin{aligned}
&\frac{1}{2 i} \oint_{\Gamma}  dz S_1'\left( z \right) = -\frac{1}{8 i} \oint_{\Gamma}  dz \frac{d}{dz} \mathrm{ln}[Q(z)] \\
&= -\left. \frac{1}{8 i} \mathrm{ln}\mathrm{\ }Q(z)\right\rvert_{\substack{\mathrm{evaluated\ once} \\ {\mathrm{around\ contour\ } \Gamma }}} = - \frac{1}{8 i} \left(2 \cdot 2\pi i \right) = -\frac{\pi}{2},
\end{aligned}
\end{equation}
where evaluating the logarithmic function around the contour $\Gamma$, enclosing the two turning points of $Q(z)$ yields $4 \pi i$.

This total contribution of $-\pi/2$ on the LHS of Eq.~\eqref{eq:Dunham1} correctly accounts for the zero-point energy of the simple harmonic oscillator. The series in Eq.~\eqref{eq:Dunham1}, truncated at the first order, is the Bohr-Sommerfeld quantization relation \cite{berry1972}. It has been considered as an exceptional case that all other higher order terms for the simple harmonic oscillator turn out to be zeros. However, in general, this is not the case. Fr\"{o}man and Fr\"{o}man \cite{froman1965} have shown that all \textit{other} higher \textit{odd}-order terms in the WKB series, Eq.~\eqref{eq:Dunham1}, can be written as exact derivatives, regardless of the type of potential, which upon contour integrating yield zeros. Setting $\epsilon=1$, we can, therefore, rewrite Eq.~\eqref{eq:Dunham1} as
\begin{equation} \label{eq:Dunham2}
\frac{1}{2 i} \oint_{\Gamma} \sum_{n=0}^{\infty} S_{2n}'\left( z \right) dz = \left(K +\frac{1}{2}\right) \pi, \qquad (K = 0,1,2,...).
\end{equation}

Several attempts have been undertaken in the past \cite{bender1977, romanovski2000} to infer the general expression for $S'_{2n}$ with expectations of summing up the series afterwards.  It, however, has turned out to be, heretofore, insurmountable. It is the objective of this work to present such a summation in an exact manner for an arbitrary potential with two turning points. (Note that such a route has been possible only for a very few special kinds of potentials, for e.g., the Eckart and the Morse potentials \cite{bender1977, romanovski2000}). Next, we outline our method of summing up the WKB series up to all-orders and interpret its physical meaning thereafter.

Let us recast Eq.~\eqref{eq:Dunham2} as
\begin{equation} \label{eq:Bohrcorrection2}
\frac{1}{2 i}  \oint_{\Gamma} S_0'\left( z \right) \cdot \sum_{n=0}^{\infty} T_{2n}\left( z \right)  dz = \left(K +\frac{1}{2}\right) \pi,
\end{equation}
where $T_{2n}(z) = S_{2n}'\left( z \right)/S_0'\left( z \right)$ and the summation over $T_{2n}$ will be achieved below. Introduce an economical notation  $L(z) \equiv 1/S_0'(z)$. ($L(z)$ can be regarded as having the dimension of length, found using Eqs.~\eqref{eq:recur} and \eqref{eq:WKB_diff}, and so does $D^{-1}$ where $D \equiv d/dz$.) Dividing both sides of Eq.~\eqref{eq:recur2} by $\left(S_0'\right)^2$ to rewrite it in these new notations of $T, D,$ and $L$, we find,
\begin{align} 
&2 T_{n} + \sum_{j = 1}^{n-1} T_{j} T_{n-j} + L^2 \frac{d}{dz} \left[ \frac{S_{n-1}'}{S_0'}\cdot \frac{1}{L} \right]= 0,\\  \label{eq:recur3}
&2 T_{n} = - \sum_{j = 1}^{n-1} T_{j} T_{n-j} - LDT_{n-1} + T_{n-1} DL.
\end{align} 

\section{Pattern Searching Campaign}
Note $T_0=1$ and $T_1=DL/2$. This allows to cast Eq.~\eqref{eq:recur3} finally in a neat way as
\begin{align}  \label{eq:recur4}
& T_{n} = T_{n-1} T_1 - \frac{1}{2}\sum_{j = 1}^{n-1} T_{j} T_{n-j} - \frac{1}{2}LDT_{n-1}, \qquad \hspace{0.3cm} (n \geq 2) .
\end{align} 

We provide below the expressions for $T_n$.\footnote{It is worth highlighting that $T_{n}$ is a dimensionless function as it has exactly the same number of $D$'s and $L$'s; see Ref.~\cite{appendix:equalnoofLandD}.} Notice the appearance of $L$ below,
\begin{subequations}
\begin{alignat}{3} \label{eq:T2}
T_2 &= &&\textrm{\ \ \ }\frac{T_1^2}{2} \quad &&- L \times \left[\frac{DT_1}{2}\right], \\
T_3 &=  && \quad &&-L\times \left[\frac{DT_2}{2}\right],\\
T_4 &=  &&-\frac{T_1^4}{8} \quad &&- L \times \left[\frac{DT_3}{2} - \frac{T_1^2\cdot DT_1}{4} + \frac{LDT_1 \cdot DT_1}{8} \right],\\
T_5 &=  && \quad &&-L \times \left[ \frac{DT_4}{2} - \frac{T_2 \cdot DT_2}{2} \right],\\
T_6 &= &&\textrm{\ \ \ }\frac{T_1^6}{16} \quad &&- L\times \left[...\right],
\end{alignat}
\end{subequations}
where ellipsis with a square bracket, $\left[...\right]$, represents a collection of functions of lower order $T_n$. Note that all terms of odd order (in $n$) of $T_n$ necessarily begin with $L$ because these expressions when substituted in Eq.~\eqref{eq:Bohrcorrection2} yield cancellation of such $L$ with $L$ in the denominator of $S_0'=1/L$. The remaining part of the integrand can be shown to be (the sum of) the product of exact derivatives or expressions that can be changed into them \cite{appendixB}. Such integrand with the product of exact derivatives is trivially zero upon contour integrating \cite{appendixC} because they are single-valued functions for the defined contour path (no logarithmic derivatives are involved here as $L$ in the denominator of $S_0'=1/L$ has been cancelled out). This is tantamount to stating that all odd-order terms of $T_n$ contribute to the wavefunction's amplitude (i.e., they do not play role in the quantization condition \cite{bender1977}) and $T_n$'s even-order terms modulate the phase of the wavefunction -- thus the quantization condition involves thereof. To reiterate, for $n \geq 1$, $T_{\mathrm{odd\ order}} = T_{2n+1}= -L \times \left[...\right]$ and $T_{\mathrm{even\ order}} = T_{2n} = ...T_1^{2n} - L \times \left[...\right]$. After explicitly computing higher order terms (e.g., $T_8 = 5T_1^8/128 - L  \times \left[...\right]$), we recognize a completely unexpected but instructive pattern, based upon which we propose the following hypothesis and prove it subsequently.\footnote{We are very grateful to Michael V. Berry for questions that prompted us to propose this hypothesis.}

\section{Inductive Hypothesis}
Proposition:
\textit{For any $n \in \mathbb {N}$,}

\begin{equation} \label{eq:proposition}
\mathrm{P}(n)\mathrm{:\ } T_{2n} = \binom{n-\frac{3}{2}}{-\frac{3}{2}}  \left(\frac{T_1}{i}\right)^{2n} - L \times \left[...\right],
\end{equation}
\textit{where $ \mathrm{\binom{n-\frac{3}{2}}{-\frac{3}{2}}} $ is the binomial coefficient and $i^2=-1$.}

This statement is not challenging to prove using the principle of mathematical induction (see section A in the Appendix for the proof).

%%%%%%%%%%%%%%%%

Although not immediately obvious, the inductive hypothesis, presented in the precise form as above, has paramount consequence. In the proposition, in Eq.~\eqref{eq:proposition}, the first term of $T_{2n}$ on the right hand side lacks $L$ in front of it unlike the second term and hence, when substituted in Eq.~\eqref{eq:Bohrcorrection2}, it yields a function that does not vanish upon doing the contour integration (as terms like $\oint_\Gamma dz S_0' \cdot T_1^{2n} \sim \oint_\Gamma dz (DL)^{2n}/L$ contribute to a logarithmic derivative of $L$ and thus the contour encloses poles of a logarithmic function, which are the zeros of $Q(z)$). In contrast thereof, the second term of $T_{2n}$ that begins with $L$, written as $L \times \left[...\right]$, contributes exactly zero upon contour integrating as the cancellation of this $L$ with $S_0'=1/L$ in Eq.~\eqref{eq:Bohrcorrection2} modifies the WKB integrand into (a sum of) the product of exact derivatives (importantly, without any logarithmic derivative) \cite{appendixB}. Resulting product of exact derivatives, lacking logarithmic term, amount to zero upon contour integrating; the reasoning here can follow the same as was aforementioned for the odd-order terms in the WKB series \cite{appendixC, appendixD}. Thus this campaign of searching terms which begin with $L$ and others which don't is unexpectedly helpful. We shall, therefore, deal with only the power series in $T_1$ in Eq.~\eqref{eq:Bohrcorrection2}. By straightforward summation of this special series up to all-orders in perturbation theory, we obtain a closed-form expression for Eq.~\eqref{eq:Bohrcorrection2} as presented below in Eq.~\eqref{eq:summedupWKB}.

\begin{widetext}
\begin{equation}  \label{eq:summedupWKB}
\begin{aligned}
\hspace{-0.2cm}\left(K +\frac{1}{2}\right) \pi
= \frac{1}{2 i} \oint_{\Gamma} \frac{dz}{L} \left[1+ \sum_{n=1,2,..}^{\infty} \binom{n-\frac{3}{2}}{-\frac{3}{2}} \left(\frac{\dl}{2 i}\right)^{2n} \right] 
=\frac{1}{2 i}\oint_{\Gamma}   \frac{dz}{L\left( z \right)} \sqrt{1+\left(\frac{DL}{2}\right)^2} &= \frac{1}{2} \oint_{\Gamma} k(z)  \cdot \sqrt{1+ \left(\frac{S_{1}'}{S_{0}'}\right)^{2}}  dz\\
&\hspace{-0.5cm}=\frac{1}{2} \oint_{\Gamma} k(z) \cdot \sqrt{1 -\left(\frac{1}{2} \frac{d (k^{-1})}{dz}\right)^{2}} dz. 
\end{aligned}
\end{equation}
\end{widetext}

We emphasize that our summation in Eq.~\eqref{eq:summedupWKB} involves a power series, which is also the case in all the special problems for which the WKB series has been summed up exactly \cite{romanovski2000, salasnich1997}. We believe this to be the reason why the first few terms of the WKB expansion often approximate the correct eigenvalues despite the full series being divergent (this is, in part, also an answer to why the WKB expansion is asymptotic).
% Further, we find that, on Taylor-expanding the final Eq.~\eqref{eq:summedupWKB}, we obtain effective contributions, found by Romanovski and Robnik \cite{romanovski2000}, to the WKB integral to all-orders of expansion for their special potentials.

We are also able to physically interpret the expressions involved on the last line of Eq.~\eqref{eq:summedupWKB} using the notion of geometric optics; however, the equation in its entirety is non-trivial to elucidate, possibly owing to the quantum effects embodied in all-orders (in perturbation theory).

\section{Geometric-Optical Meaning}
We borrow here the illustration by Bremmer \cite{bremmer1951} where the author demonstrates, \textit{mutatis mutandis}, the similarity of each order of WKB series (consider its $n^{\mathrm{th}}$ order) to the transmitted waves in an infinitesimally-discretized inhomogeneous medium that undergo $n$-number of reflections. At each reflection, the waves change their direction by $180^{{\circ}}$. Thus a wave that begins from a point far to the left gets continually transmitted to the right while suffering reflection at each discretized boundary (and consider for now only the directly transmitted waves -- not doubly or quadruply or even number of multiply reflected ones that also can eventually transmit to the right). Such resulting wavefunction at the rightmost end yields the $1^{\mathrm{st}}$ order WKB approximation \cite{bremmer1951}. Each of the above-mentioned reflected waves can undergo further reflection(s) and keep continually transmitting to the right. The more the number of reflections they suffer before they arrive to the rightmost end, the higher they belong to in the order of WKB expansion. Referring readers to the original paper \cite{bremmer1951} for additional interesting details, we present now heuristically how Bremmer arrives at the reflection coefficient of a layer of width $1/k(z)$. The well-known reflection coefficient in one-dimension is (Bremmer \cite{bremmer1951} uses $R$ notation, which we shall reserve for the reflectivity to avoid potential confusions),
\begin{equation}
r(z) = \frac{k_s - k_{s+1}}{k_s + k_{s+1}} \approx \frac{k_s - k_{s+1}}{2 k_s}  \approx-\frac{dk/d\xi}{2k}, 
\end{equation}
where $s$ is the layer-number in the infinitesimally discretized inhomogeneous medium (within which $k_s$ remains constant) and $\xi$ is proportional to the number of wavelengths over which the wavelength (or $k$) changes appreciably (from $k_s$ to $k_{s+1}$). Following Bremmer \cite{bremmer1951}, $d\xi = k(z) dz$.
Therefore, the transmissivity of a layer of width $1/k(z)$ is
\begin{equation}
    T(z) = 1-r^2(z) = 1-\left(-\frac{dk/d\xi}{2k}\right)^2=1 -\left(\frac{1}{2} \frac{d (k^{-1})}{dz}\right)^{2},
\end{equation}
which is exactly what appears on the last line of Eq.~\eqref{eq:summedupWKB}. 

Note that the wavenumber, obtained via resummation of WKB series to all-orders (i.e., integrand in Eq.~\eqref{eq:summedupWKB}), vanishes even before we reach the classical turning points. Interestingly, for \textit{all} potentials, it vanishes exactly at the locations where $p_{\mathrm{cl}}(x) \cdot x = \hslash/2$ where $p_{\mathrm{cl}} = \sqrt{2m[E-V(x)]}$ (cf. with the Heisenberg uncertainty principle). 

\section{Applications}
Our novel closed-form quantization condition yields exact energy eigenvalues to potentials with two turning points. We demonstrate the efficacy of our formula in an example below and present other cases like simple harmonic oscillator, $3$-dimensional harmonic oscillator, Coulomb potential, Eckart potential, and Morse potential in section D of the Appendix. We also find in cases of $3$-dimensional spherically symmetric potentials, the Langer-correction factor \cite{langer1937, koike2009} appears naturally upon performing the contour integration of our formula that resums all-orders perturbative effects. This corroborates the previous claim that the Langer-modification comes from the higher order corrections in the WKB series \cite{salasnich1997}. 

Consider an asymmetric Rosen-Morse potential, $V(x)$, with $\hslash^2=2m$:
$V(x)=-U_0 \textrm{sech}^2\left( x/a \right) +U_1\tanh\left( x/a \right)$. Let $z$ be a complex variable such that $z=\tanh\left( x/a\right)$. Then, $z_a$ and $z_b$ are the two classical turning points, satisfying $z_a+z_b = -U_1/U_0$; $z_a z_b = -\left(E+U_0\right)/U_0$. Using Eq.~\eqref{eq:quantpot},

\begin{equation}  \label{eq:example1}
\frac{1}{2 i}\oint_{\Gamma}  \frac{\sqrt{16 \left[ V(z)-E \right]^{3}+\left[V'(z)\right]^{2} }}{4 \left[ V(z)-E \right]} dz = \left(K +\frac{1}{2}\right)\pi,
\end{equation}
where $K=0,1,2,...$ represent different energy levels.
The poles of the integrand are at $z=z_a,z_b,1,-1,\textrm{ and } \infty$. So, calculating residue at each pole with a proper principal value yields
\begin{align}
\begin{split}
-\frac{1}{4}+\frac{1}{4}  &{-} \frac{a\sqrt{(z_a-1)(z_b-1)U_0}}{2} {-} \frac{a\sqrt{(z_a+1)(z_b+1)U_0}}{2}\\
&\hspace{2cm}+ \frac{\sqrt{1+4a^2 U_0}}{2} = K+\frac{1}{2},    
\end{split}
\end{align}
\begin{equation}
\begin{split}
%{+}
\therefore \frac{1}{2}\sqrt{\frac{-E-U_1}{U_0}} {+} \frac{1}{2}\sqrt{\frac{-E+U_1}{U_0}} = -\frac{1}{a\sqrt{U_0}} \left( K+\frac{1}{2} \right) \\
+ \frac{\sqrt{1+4a^2U_0 }}{2a\sqrt{U_0}},
\end{split}
\end{equation} 
which agrees with Ma \& Xu  \cite{maxu2005} and the individual terms directly manifest from the residues at poles whose locations are precisely predicted by Eq.~\eqref{eq:example1}.

At last, we remark that, for reasons not fully understood, the proposed quantization relation works only for two turning point problems. Nevertheless,
understanding the nature of the exactly quantized action in two turning point problems, considered here, is likely to benefit the extension of our geometric-optically interpretable equation to problems with multiple turning points. Such interesting extensions will be investigated in a future study. Our proposed quantization condition might also engender rethinking of quantization in higher dimensions in terms of geometry \cite{stone2005}.

\section{Conclusion}
This article presents an exact closed-form quantization relation by summing up the WKB series to all-orders in perturbation theory for arbitrary one-dimensional potentials having two turning points. The new resummation procedure utilized herein reveals an unexpectedly simple pattern in the general term of the WKB series, leading to an inductive hypothesis, which we are able to prove by the principle of mathematical induction. The presented formula is then physically interpreted as the action of a wave with wavenumber corrected by a factor related to the wave transmissivity. Unit transmissivity recovers the Bohr-Sommerfeld quantization relation. This closed-form expression for the quantization might also be useful in problems with more than two turning points where non-perturbative effects that give rise to tunneling phenomena come into play, i.e., spectral curves with non-zero genus \cite{basar2017}. For such problems, some of the neglected terms in the series resummation appear to be necessary. In the light of resurgent perturbative/nonperturbative relations \cite{dunne2014, Gahramanov2016, basar2017}, collecting such terms seem interesting and further investigation is merited. This will, however, be left for the future as it is beyond the objective of the present article. Arguably the most important advancement through this work is the discovery of an elegant and physically-interpretable equation emerging from a myriad of complicated terms in the WKB series that become increasingly unmanageable at each higher order of the series. It is gratifying to find that the spectral problem with genus-0 spectral curve (i.e., with two turning points and no tunneling phenomenon) can be reduced to an exact equation as simple and economical as Eq.~\eqref{eq:quantpot}. Analyzing this equation might lead to a deeper understanding of quantum geometry \cite{basar2017}.

\begin{acknowledgments}
I am pleased to thank Dhrubaditya Mitra for enkindling interest in this work and for offering discussions and guidance; it is difficult to imagine NORDITA (Sweden) without his friendly teaching style and advocacy for perturbation methods. I am also indebted to Michael V. Berry and Carl M. Bender for their appraisals of this work that greatly improved the manuscript. Thanks are also due to Paul W. Terry, MJ Pueschel, and Dibyendu Nandi for their valuable feedback and suggestions. Correspondences with Mithat \"{U}nsal and Luca Salasnich helped to find connections of this work with others. I also acknowledge support from the Van Vleck Fellowship in Physics at UW-Madison and the conducive environment herein.
\end{acknowledgments}

% \pagebreak
% \newpage
% \newpage

%%%%%%%%%%%%%%%%%%%%%%%%%%%%%%%%%%%%%%%%
\begin{widetext}
\vspace{50cm}
\section{Appendix}

\subsection{Proof by the principle of mathematical induction}

Below, we present the proof for the proposition put forth for $T_{2n}$ in Eq.~\eqref{eq:proposition}. The statement, P($n$), is trivially satisfied for $n=1$ [cf. Eq.~\eqref{eq:proposition} with Eq.~\eqref{eq:T2}]. Now, we assume it to be valid for an arbitrary $n$ and prove that it implies the proposition, in Eq.~\eqref{eq:proposition}, is true for $n+1$ as well [i.e., P$(n+1)$ is true].  We begin with the WKB recurrence relation, Eq.~\eqref{eq:recur4},
\begin{subequations}
\begin{align} 
 T_{2(n+1)} 
&= - \frac{1}{2}\sum_{j = 2}^{2(n+1)-2} T_{j} T_{2(n+1)-j} - \frac{LDT_{2(n+1)-1}}{2}\\
&= -\sum_{\substack{j = 2,4,... \\ \mathrm{\textbf{even}}}}^{2n} \frac{T_{j} T_{2n+2-j}}{2}  - \sum_{\substack{j = 3,5,...\\ \mathrm{\textbf{odd}}}}^{2n-1} \frac{T_{j} T_{2n+2-j}}{2} - \frac{LDT_{2n+1}}{2}\\
&= \left\{-\sum_{j/2 = 1,2,...}^{n} \binom{\frac{j}{2}-\frac{3}{2}}{-\frac{3}{2}} \binom{n+1-\frac{j}{2}}{-\frac{3}{2}} \left(\frac{T_{1}}{i}\right)^{2n+2} - L \times \left[...\right] \right\} - \sum_{\substack{j = 3,5,...\\ \mathrm{\textbf{odd}}}}^{2n-1} \frac{T_{j} T_{2n+2-j}}{2} - \frac{LDT_{2n+1}}{2}\\
&= -\sum_{j/2 = 1,2,...}^{n} \binom{\frac{j}{2}-\frac{3}{2}}{-\frac{3}{2}} \binom{n+1-\frac{j}{2}}{-\frac{3}{2}} \left(\frac{T_{1}}{i}\right)^{2n+2} - L \times \left[...\right] -L \times \left[...\right] - \frac{LDT_{2n+1}}{2} \hspace{0.5cm} \left[ \because T_{\mathrm{odd}} \mathrm{\ begins\ with\ } L \right] \\
&= \binom{n+1-\frac{3}{2}}{-\frac{3}{2}}  \left(\frac{T_1}{i}\right)^{2(n+1)} - L \times \left[...\right] \QEDA
\end{align} 
\end{subequations}
This proves the proposition.

\subsection{$T_n$ beginning with $L$ convertible to a product of exact derivatives}
Here, we detail the recipe of casting any expression in $T_n$ that begins with $L$ into a product of exact derivatives (without a logarithmic derivative) \cite{appendixB}. Consider the term in $T_n$ that begins with $L$: 
\begin{equation} \label{eq:appB1}
    \begin{aligned}
    \oint_\Gamma dz S_0' T_n  = \oint_\Gamma dz S_0' L \times \left[...\right] &= \oint_\Gamma dz \times \frac{1}{L} \times L \times \left[...\right] \\
    &= \oint_\Gamma dz \times \frac{1}{L} \times L \times \left[\left(D^{p_1}L^{q_1}D^{p_2}L^{q_2}...\right) \times ... \times \left(D^{p_{j-1}}L^{q_{j-1}}D^{p_{j}}L^{q_{j}}...\right)\right]\\
    &= \oint_\Gamma dz  \left[\left(D^{p_1}L^{q_1}D^{p_2}L^{q_2}...\right) \times ... \times \left(D^{p_{j-1}}L^{q_{j-1}}D^{p_{j}}L^{q_{j}}...\right)\right],
    \end{aligned}
\end{equation}
where $D^{p_1}L^{q_1}D^{p_2}L^{q_2}...$ represents $D^{p_1}\left(L^{q_1}D^{p_2}\left(L^{q_2}...\right)\right)$; $p_1, q_1, p_2, q_2, ..., p_j, q_j, ...$ are all integers in between (and including) $0$ and $n$, satisfying the constraints: $p_1+p_2+...+p_j+... = n$ (for $n$ number of $D$'s) and $q_1+q_2+...+q_j+... = n-1$ (for $n-1$ number of $L$'s); and one extra $L$ lies in the very beginning of $T_n$, which has gotten cancelled with $S_0'=1/L$. This makes $T_n$ to have $n$ number of $D$'s and $L$'s, as argued in Ref.~\cite{appendix:equalnoofLandD}. If all $p_1, p_2, ..., p_j, ...$ are equal to or greater than $1$, this integrand is already a product of exact derivatives (without a logarithmic derivative as $S_0'=1/L$ has already been cancelled with $L$ of $T_n$). If any of $p_1,p_2, ...,p_j, ..$ is zero, the integrand in Eq.~\eqref{eq:appB1}, using chain rule, becomes (say $p_1=0$): \begin{equation}
    \begin{aligned}
    \mathrm{integrand\ } &= \left(D^{0}L^{q_1}D^{p_2}L^{q_2}...\right) \times ... \times \left(D^{p_{j-1}}L^{q_{j-1}}D^{p_{j}}L^{q_{j}}...\right) \\
    &= \hspace{0.5cm}\left(L^{q_1}D^{p_2}L^{q_2}...\right) \times ... \times \left(D^{p_{j-1}}L^{q_{j-1}}D^{p_{j}}L^{q_{j}}...\right)\\
    &= D\left[\left(L^{q_1}D^{p_2}L^{q_2}...\right) \times ... \times \left(D^{p_{j-1}}L^{q_{j-1}}D^{p_{j}}L^{q_{j}}...\right)\right] \bm{-} D\left[\left(L^{q_1}D^{p_2}L^{q_2}...\right) \times ...  \right] \times \left(D^{p_{j-1}-1}L^{q_{j-1}}D^{p_{j}}L^{q_{j}}...\right),
    \end{aligned}
\end{equation}
thus turning the integrand into (a product of) exact derivatives. This process can be repeated if $p_{j-1}-1$ also happens to be $0$ (and if several $L$ multiplies each other, its exponent can be raised to abridge this procedure). It should be emphasized that it is guaranteed to find such a transformation as there are exactly the same number of $D$'s and $L$'s in any $T_n$ \cite{appendix:equalnoofLandD}. In this regard, searching a general expression for $T_{2n}$ as stated in the proposition in Eq.~\eqref{eq:proposition} is surprisingly helpful.

\subsection{Product of exact derivatives yields zero}
We demonstrate here that the product of exact derivatives (without a logarithmic function) amounts to zero on contour integrating, expounding on Ref.\cite{appendixC}. Let us consider two functions $f$ and $g$ (without a logarithm). Then, $\oint_\Gamma dz ~ df/dz  = 0$.
% \begin{align}
% \oint_\Gamma dz \frac{df}{dz}  &= 0
% \end{align}
Now, using the Cauchy integral formula to represent the exact derivatives and thereafter employing the partial fraction decomposition,
\begin{subequations}
\begin{align}
\oint_\Gamma dz \frac{df}{dz} \frac{dg}{dz}  &= \oint_\Gamma dz \frac{df}{dz} \frac{dg}{dz} \\
&= \oint_\Gamma dz \left\{\frac{1}{2\pi i} \oint_{\gamma_1} \frac{f(u)}{\left(u-z\right)^{1+1}} du \right\} \left\{\frac{1}{2\pi i} \oint_{\gamma_2} \frac{g(v)}{\left(v-z\right)^{1+1}} dv \right\} \hspace{1cm} [\gamma_1\ \& \ \gamma_2\ \mathrm{enclose\ the\ pole\ at\ } z]\\
&= -\frac{1}{4 \pi^2} \oint_{\gamma_1} \oint_{\gamma_2} du dv f(u) g(v)  \oint_\Gamma dz {\left[\frac{1}{(u-z)(v-z)}\right]}^2\\
&= -\frac{1}{4 \pi^2} \oint_{\gamma_1} \oint_{\gamma_2} du dv f(u) g(v)  \oint_\Gamma dz {\left[\frac{1}{v-u} \left(\frac{1}{u-z}-\frac{1}{v-z} \right)\right]}^2\\
&= -\frac{1}{4 \pi^2} \oint_{\gamma_1} \oint_{\gamma_2} du dv \frac{f(u) g(v)}{\left(v-u\right)^2}  \oint_\Gamma dz { \left(\frac{1}{u-z}-\frac{1}{v-z} \right)}^2\\
&= -\frac{1}{4 \pi^2} \oint_{\gamma_1} \oint_{\gamma_2} du dv \frac{f(u) g(v)}{\left(v-u\right)^2} \left[ \oint_\Gamma dz \frac{1}{\left(u-z\right)^2} -2\oint_\Gamma dz \frac{1}{\left(u-z\right)\left(v-z\right)} + \oint_\Gamma dz \frac{1}{\left(v-z\right)^2} \right]\\
&= -\frac{1}{4 \pi^2} \oint_{\gamma_1} \oint_{\gamma_2} du dv \frac{f(u) g(v)}{\left(v-u\right)^2} \left[ 0 -2\oint_\Gamma dz \frac{1}{\left(u-z\right)\left(v-z\right)} + 0 \right]\\
&= \frac{1}{2 \pi^2} \oint_{\gamma_1} \oint_{\gamma_2} du dv \frac{f(u) g(v)}{\left(v-u\right)^2}  \oint_\Gamma dz  \frac{1}{v-u} \left(\frac{1}{u-z}- \frac{1}{v-z} \right)\\
&= \frac{1}{2 \pi^2} \oint_{\gamma_1}  \oint_{\gamma_2} du dv \frac{f(u) g(v)}{\left(v-u\right)^3} \left[ \oint_\Gamma dz \frac{1}{u-z}-  \oint_\Gamma dz \frac{1}{v-z} \right]\\
&= \frac{1}{2 \pi^2} \oint_{\gamma_1}  \oint_{\gamma_2} du dv \frac{f(u) g(v)}{\left(v-u\right)^3} \left[- 2 \pi i+ 2 \pi i\right]\\
&=0.
\end{align}
\end{subequations}
It is straightforward to prove, in the similar manner, that the product of three (or more) exact derivatives ((without a logarithmic function) under contour integration is also zero. We are thus left with a power series in $\left(T_1\right)^{2n}$, which we sum up in Eq.~\eqref{eq:summedupWKB}.

\end{widetext}
\subsection{More Examples}
Let us now test the validity of our novel formula. We find that this formula gives the exact energy eigenvalues $E$ for all the following potentials and many more which are not listed here (but all having exactly two turning points). We choose, without the loss of generality, $\hslash^2 = 2m$ in all of the following calculations.

\subsection*{I. Simple Harmonic Oscillator}
Even though we already know that the leading-order WKB is exact for simple harmonic oscillator, it is desirable to test if the extra factor in the integrand of Eq.~\eqref{eq:example1} would cause any deviation from the correct eigenvalues. Consider,
\begin{subequations}
\begin{align}
V(z)&=z^2,\\
V-E &=z^2-E,\\
V'(z) &=2 z .
\end{align}
\end{subequations}

From Eq.~\eqref{eq:example1},
\begin{equation} 
\frac{1}{2 i}\oint  \frac{\sqrt{16 \left( V-E \right)^{3}+\left(V'\right)^{2} }}{4 \left( V-E \right)} dz = \left(K +\frac{1}{2}\right)\pi,
\end{equation}

which leads to
\begin{equation} 
\frac{1}{2 i}\oint  \frac{\sqrt{16 \left( z^2-E \right)^{3}+\left( 2z \right)^{2} }}{4 \left( z^2-E \right)} dz = \left(K +\frac{1}{2}\right)\pi. \hspace{0.6cm} 
\end{equation}
The poles of the integrand are at $z=\sqrt{E},-\sqrt{E},\textrm{ and } \infty$. So, calculating residues at each of the poles respectively,
\begin{align}
 \frac{1}{4}-\frac{1}{4} + \frac{E}{2} &= K+\frac{1}{2},\\
\therefore E &=2 \left( K+\frac{1}{2}\right).
\end{align}

\subsection*{II. $3$-D harmonic oscillator  }
Consider the potential,
\begin{subequations}
\begin{align}
V(r)&=r^2+\frac{b}{r^2}+\frac{l(l+1)}{r^2},\\
V-E &= r^2+\frac{b}{r^2}+\frac{l(l+1)}{r^2}-E.
\end{align}
\end{subequations}

Let $u=r^2$; $u' = 2r= 2\sqrt{u}$.
\begin{subequations}
\begin{align}
\therefore V-E &= \frac{1}{u} \left[ u^2-E u + \left\{b+l\left(l+1\right)  \right\}  \right]\\
&= \frac{1}{u} \left(u-u_a  \right)\left(u-u_b  \right),
\end{align}
\end{subequations}
with 
\begin{subequations}
\begin{align}
u_{a,b} &= \frac{E}{2}\pm \sqrt{\left(\frac{E}{2}\right)^2 -\left\{b+l\left(l+1\right) \right\} },\\
u_a+u_b &= E,\\
u_a u_b &= b+l\left(l+1\right).
\end{align}
\end{subequations}
Now,
\begin{subequations}
\begin{align}
V' &= \frac{dV}{du}u'\\
&= \left[ \frac{2u-u_a-u_b}{u} -\frac{(u-u_a)(u-u_b)}{u^2}\right] 2\sqrt{u}.
\end{align}
\end{subequations}

Using Eq.~\eqref{eq:example1},
\begin{equation} 
\frac{1}{2 i}\oint  \frac{\sqrt{16 \left( V-E \right)^{3}+\left(V'\right)^{2} }}{4 \left( V-E \right)} dz = \left(K +\frac{1}{2}\right)\pi.
\end{equation}
The poles of the integrand are at $u=u_a,u_b,0,\textrm{ and } \infty$. So, calculating residues at each of the poles respectively,

\begin{align}
 \frac{1}{4}-\frac{1}{4} - \frac{1}{4}&\sqrt{1+4u_a u_b}+ \frac{u_a+u_b}{4} = K+\frac{1}{2},\\
\therefore E &=2 \left[ 2K+1+\sqrt{{\color{red}\bm{\left(l+\frac{1}{2}\right)}^2}+b}\right].
\end{align}

This solution agrees with Rosenzweig \& Krieger \cite{Rosenzweig1968}. \\

Note that the correct Langer correction factor, shown in the bold typeset, emerges naturally from our all-orders resummed WKB series, which Langer proposed to replace $l(l+1)$ by $(l+1/2)^2$ to obtain the correct eigenvalues.

\subsection*{III. Coulomb potential}
Now, assume the Coulomb potential,
\begin{subequations}
\begin{align}
V(r)&=-\frac{V_0}{r}+\frac{b}{r^2}+\frac{l(l+1)}{r^2},\\
V-E &= \frac{1}{r^2}\left[-V_0 r + b+ l(l+1)-Er^2 \right]\\
&= \frac{-E}{r^2}\left[r^2+\frac{V_0}{E}r-\frac{b+l(l+1)}{E}  \right]\\
&= \frac{-E}{r^2}\left(r-r_a\right)\left(r-r_b\right),
\end{align}
\end{subequations}
with 
\begin{subequations}
\begin{align}
r_{a,b} &= -\frac{V_0}{2E}\pm \sqrt{\left(\frac{V_0}{2E}\right)^2 +\frac{b+l\left(l+1\right) }{E} },\\
r_a+r_b &= \frac{-V_0}{E},\\
r_a r_b &= -\frac{b+l\left(l+1\right)}{E}.
\end{align}
\end{subequations}
Now,
\begin{align}
V' &= \frac{2E(r-r_a)(r-r_b)}{r^3} - \frac{E}{r^2}\left(2r-r_a-r_b\right).
\end{align}

We use Eq.~\eqref{eq:example1} below,
\begin{equation} 
\frac{1}{2 i}\oint  \frac{\sqrt{16 \left( V-E \right)^{3}+\left(V'\right)^{2} }}{4 \left( V-E \right)} dz = \left(K +\frac{1}{2}\right)\pi.
\end{equation}
The poles of the integrand are at $r=r_a,r_b,0,\textrm{ and } \infty$. So, calculating residues at each of the poles with proper principal value,

\begin{align}
- \frac{1}{4}+\frac{1}{4} {-} \frac{1}{2}&\sqrt{1-4E r_a r_b}+ \frac{\sqrt{-E}}{2}\left(r_a+r_b\right) = K+\frac{1}{2},\\
\therefore E &= \frac{-V_0^2}{4 \left[ K+1/2 {+}\sqrt{b+{\color{red}\bm{\left(  l+\frac{1}{2}\right)}^2}}  \right]^2}.
\end{align}

This solution agrees with Rosenzweig \& Krieger \cite{Rosenzweig1968}.\\

Notice again that the correct Langer correction factor, shown in the bold typeset, emerges naturally from our all-orders resummed WKB series.

\subsection*{IV. Eckart potential}
Let us consider the Eckart potential,
\begin{align}
V(x)&=\frac{-\lambda e^{-\alpha x}}{1-e^{-\alpha x}}+\frac{b e^{-\alpha x}}{\left(1-e^{-\alpha x}\right)^2}.
\end{align}

Using a transformation, $u=e^{\alpha x}-1$; $u' = \alpha(u+1)$; we write,
\begin{subequations}
\begin{align}
V-E &= \frac{-\lambda}{u} +\frac{b(u+1)}{u^2}-E\\
&= \frac{-E}{u^2}\left[u^2+\frac{\lambda-b}{E}u -\frac{b}{E}  \right]\\
&= \frac{-E}{u^2}(u-u_a)(u-u_b),
\end{align}
\end{subequations}
with 
\begin{subequations}
\begin{align}
u_{a,b} &= \frac{b-\lambda}{2E}\pm \sqrt{\left(\frac{b-\lambda}{2E}\right)^2 +\frac{b}{E} },\\
u_a+u_b &= \frac{b-\lambda}{E},\\
u_a u_b &= -\frac{b}{E}.
\end{align}
\end{subequations}
Now,
\begin{subequations}
\begin{align}
V' &= \frac{dV}{du}u'\\
&= \left[\frac{2E}{u^3}(u-u_a)(u-u_b) -  \frac{E}{u^2}(2u-u_a-u_b) \right] \alpha (u+1).
\end{align}
\end{subequations}

Using Eq.~\eqref{eq:example1},
\begin{equation} 
\frac{1}{2 i}\oint  \frac{\sqrt{16 \left( V-E \right)^{3}+\left(V'\right)^{2} }}{4 \left( V-E \right)} dz = \left(K +\frac{1}{2}\right)\pi.
\end{equation}
The poles of the integrand are at $u=u_a,u_b,0,-1,\textrm{ and } \infty$. So, calculating residues at each of the poles with proper principal value,

\begin{eqnarray}
\begin{split}
 -\frac{1}{4}+\frac{1}{4} {-} \frac{1}{2\alpha}\sqrt{\alpha^2-4E u_a u_b}+& \frac{1}{\alpha}\sqrt{-E(1+u_a)(1+u_b)}\\
&{-} \frac{\sqrt{-E}}{\alpha} = K+\frac{1}{2},
 \end{split}\\
\therefore {-} \frac{1}{2}\sqrt{1+\frac{4b}{\alpha^2}}+ \frac{\sqrt{\lambda-E}}{\alpha}{-} \frac{\sqrt{-E}}{\alpha} = K+\frac{1}{2}\hspace{1cm}
\end{eqnarray}

This solution agrees with Romanovski \& Robnik \cite{romanovski2000}.

\subsection*{V. Morse potential}
Finally, consider the potential of the form,
\begin{subequations}
\begin{align}
V(x)&=Ae^{-2\alpha x}-Be^{-\alpha x},\\
V-E &= Ae^{-2\alpha x}-Be^{-\alpha x}-E.
\end{align}
\end{subequations}

Let $u=e^{\alpha x}$, $u' = \alpha u$, which leads us to,
\begin{subequations}
\begin{align}
\therefore V-E &= \frac{1}{u^2} \left[ A-Bu-Eu^2 \right]\\
&= \frac{-E}{u^2} \left[ u^2+\frac{B}{E}u -\frac{A}{E}\right]\\
&= \frac{-E}{u^2} (u-u_a)(u-u_b),
\end{align}
\end{subequations}
with 
\begin{subequations}
\begin{align}
u_{a,b} &= -\frac{B}{2E}\pm \sqrt{\left(\frac{B}{2E}\right)^2 +\frac{A}{E} },\\
u_a+u_b &= \frac{-B}{E},\\
u_a u_b &= \frac{-A}{E}.
\end{align}
\end{subequations}
Now,
\begin{subequations}
\begin{align}
V' &= \frac{dV}{du}u'\\
&= \left[\frac{2E}{u^3}(u-u_a)(u-u_b) -  \frac{E}{u^2}(2u-u_a-u_b) \right] \alpha u.
\end{align}
\end{subequations}

Employing Eq.~\eqref{eq:example1}, 
\begin{equation} 
\frac{1}{2 i}\oint  \frac{\sqrt{16 \left( V-E \right)^{3}+\left(V'\right)^{2} }}{4 \left( V-E \right)} dz = \left(K +\frac{1}{2}\right)\pi.
\end{equation}
The poles of the integrand are at $u=u_a,u_b,0,\textrm{ and } \infty$. So, calculating residues at each of the poles with proper principal value,

\begin{align}
 \frac{-1}{4}+\frac{1}{4} + \frac{1}{2 \alpha}\sqrt{\frac{-E}{u_a u_b}}(u_a+u_b) {-}\frac{\sqrt{-E}}{\alpha} &= K+\frac{1}{2},\\
\therefore\frac{B}{2\alpha \sqrt{A}} {-}\frac{\sqrt{-E}}{\alpha} = K+\frac{1}{2}.
\end{align}

This solution agrees with Romanovski \& Robnik \cite{romanovski2000}.


\begin{thebibliography}{9}
\bibitem{appendix:equalnoofLandD}
Note that $L$'s and $D$'s appear exactly $n$ number of times in any $T_n$ as it is constructed to be so in Eq.~\eqref{eq:Bohrcorrection2}. This can be confirmed via dimensional arguments (as $D$'s and $L$'s have the dimensions of each other's inverse and $T_n$ is a dimensionless quantity with $n$ number of derivatives). One can also observe this via Eq.~\eqref{eq:recur4} by noticing the appearances of $D$ and $L$, with $T_0=1$ and $T_1=DL/2$.
\bibitem{appendixB}
Any $T_n$ in Eq.~\eqref{eq:Bohrcorrection2}, by construct, has exactly $n$ number of $D$'s and $L$'s each as argued in Ref. \cite{appendix:equalnoofLandD}. This implies, non-trivially, that any expression in $T_n$ that begins with $L$ can be changed into a product of exact derivatives (without a logarithmic derivative). It is because the $L$ that $T_n$ begins with, cancels the factor $S_0'=1/L$, the latter of which is responsible for a logarithmic derivative. After this cancellation, all $L$'s appear in the numerator ($D$'s always manifest in the numerator). For more explicit details, see section B in the Appendix.
\bibitem{appendixC}
A heuristic proof can be obtained using the Cauchy integral formula (CIF) to represent each of the exact derivatives and switching the WKB contour integration's order with the contour integrations issued by the CIF. Detailed calculations are presented in section C of the Appendix.
\bibitem{appendixD}
A more rigorous analysis is likely to help extend our formula to problems with more than two turning points.
\end{thebibliography}
\end{document}